\documentclass[11pt]{article}
\usepackage{pstricks}
\usepackage{fullpage}
\usepackage{graphicx}
\usepackage{amsmath}
\usepackage{amssymb}
\usepackage{relate}

\newtheorem{lemma}{Lemma}
\newtheorem{theorem}{Theorem}
\newtheorem{corollary}{Corollary}

\newcommand{\qed}{\hfill\ensuremath{\Box}\medskip\\\noindent}
\newenvironment{proof}{\noindent\emph{Proof. }}

\newcommand{\ceil}[1]{\left\lceil{#1}\right\rceil}

\newcommand{\pqtime}[2]{\ensuremath{\left\langle{#1},{#2}\right\rangle}}

\newcommand{\Pre}{\ensuremath{\mathrm{Pre}}}
\newcommand{\prcfont}[1]{{\ensuremath{\mathsf{#1}}}}
\newcommand{\Succ}{\prcfont{Succ}}
\newcommand{\Eq}{\prcfont{Eq}}
\newcommand{\Move}{\prcfont{Move}}
\newcommand{\Close}{\prcfont{Close}}
\newcommand{\Next}{\prcfont{Next}}

\title{Fast and Compact Regular Expression Matching}

\author{
    Philip Bille
    \thanks{IT University of Copenhagen, 2300 Copenhagen S, Denmark;
        \texttt{beetle@itu.dk}. Research supported by the Danish Agency for Science, Technology, and Innovation.}
    \and
    Martin~Farach-Colton
    \thanks{Department of Computer Science, Rutgers University, Piscataway, NJ 08855, USA;
    \texttt{farach@cs.rutgers.edu}.}
}

\date{\today}

\begin{document}

\maketitle

\begin{abstract}
We study $4$ problems in string matching, namely, regular expression matching, approximate regular expression matching, string edit distance, and subsequence indexing, on a  standard word RAM model of computation that allows logarithmic-sized words to be manipulated in constant time. We show how to improve the space and/or remove a dependency on the alphabet size for each problem using either an improved tabulation technique of an existing algorithm or by combining known algorithms in a new way.
\end{abstract}

{\bf Keywords:} Regular Expression Matching; Approximate Regular Expression Matching: String Edit Distance; Subsequence Indexing; Four Russian Technique.

\section{Introduction}\label{sec:intro}

We study $4$ problems in string matching on a standard word RAM model of computation that allows logarithmic-sized words to be manipulated in constant time. This model is often called the \emph{transdichotomous model}. We show how to improve the space and/or remove a dependency on the alphabet size for each problem. Three of the results are obtained by improving the tabulation of subproblems within an existing algorithm. The idea of using tabulation to improve algorithms is often referred to as the \emph{Four Russian Technique} after Arlazarov et al.~\cite{ADKF1970} who introduced it for boolean matrix multiplication. The last result is based on a new combination of known algorithms. The problems and our results are presented below.

\paragraph{Regular Expression Matching}
Given a regular expression $R$ and a string $Q$, the \textsc{Regular
  Expression Matching} problem is to determine if $Q$ is a member of
  the language denoted by $R$. This problem occurs in several text
  processing applications, such as in editors like
  Emacs~\cite{Stallman1981} or in the \texttt{Grep}
  utilities~\cite{WM1992,Navarro2001}.  It is also used in the lexical
  analysis phase of compilers and interpreters, regular expressions
  are commonly used to match tokens for the syntax analysis phase, and
  more recently for querying and validating XML databases, see
  e.g.,~\cite{HP2001,LM2001,Murata2001,BMLMA2004}. The standard
  textbook solution to the problem, due to Thompson~\cite{Thomp1968},
  constructs a non-deterministic finite automaton (NFA) for $R$ and
  simulates it on the string $Q$. For $R$ and $Q$ of sizes $m$ and
  $n$, respectively, this algorithm uses $O(mn)$ time and $O(m)$
  space. If the NFA is converted into a deterministic finite automaton
  (DFA), the DFA needs $O(\frac{m}{w} 2^{2m}\sigma)$ words, where
  $\sigma$ is the size of the alphabet $\Sigma$ and $w$ is the word
  size.  Using clever representations of the DFA the space can be reduced to $O(\frac{m}{w}(2^m +
  \sigma))$~\cite{WM1992b,NR2004}. Efficient average case algorithms were given by Baeza-Yates and Gonnet~\cite{BYG1996}.

Normally, it is reported that the running time of traversing the DFA
is $O(n)$, but this complexity analysis ignores the word size. Since
nodes in the DFA may need $\Omega(m)$ bits to be addressed, we may need
$\Omega(m/w + 1)$ time to identify the next node in the
traversal. Therefore the running time becomes $O(mn/w + n + m)$ with a
potential exponential blowup in the space. Hence, in the
transdichotomous model, where $w$ is $\Theta(\log (n + m))$, using
worst-case exponential preprocessing time improves the query time by a
log factor.  



The fastest known algorithm is due to
Myers~\cite{Myers1992}, who showed how to achieve $O(mn/k + m2^k + (n
+ m)\log m)$ time and $O(2^km)$ space, for any $k \leq w$. In
particular, for $k = \epsilon\log n$, for constant $0 < \epsilon < 1$, this gives an algorithm using
$O(mn/\log n + (n+ m)\log m)$ time and $O(mn^{\epsilon})$ space.

In Section \ref{sec:regex}, we present an algorithm for \textsc{Regular
Expression Matching} that takes time $O(nm/k + n + m\log m)$ time and
uses $O(2^k + m)$ space, for any $k\leq w$. In particular, if we pick $k=\epsilon \log n$, for constant $0 < \epsilon < 1$, we
are (at least) as fast as the algorithm of Myers, while achieving $O(n^{\epsilon}+m)$ space.

We note that for large word sizes ($w > \log^2 n$) one of the authors has recently devised an even faster algorithm using very different ideas~\cite{Bille06}. This research was done after the work that led to the results in this paper.

\paragraph{Approximate Regular Expression Matching}
Motivated by applications in computational biology, Myers and
Miller~\cite{MM1989} studied the \textsc{Approximate Regular
  Expression Matching} problem. Here, we want to determine if
$Q$ is within \emph{edit distance} $d$ to any string in the language
given by $R$. The edit distance between two strings is the minimum
number of insertions, deletions, and substitutions needed to transform
one string into the other. Myers and Miller~\cite{MM1989} gave an
$O(mn)$ time and $O(m)$ space dynamic programming
algorithm. Subsequently, assuming as a constant sized alphabet, Wu, Manber and Myers~\cite{WMM1995} gave an $O(\frac{mn\log(d+2)}{\log n} + n + m)$ time and $O(\frac{m\sqrt{n}
  \log(d+2)}{\log n} + n + m)$ space algorithm. Recently, an exponential space solution based on DFAs for the problem has been proposed by Navarro~\cite{Navarro2004}. 
  
  In Section \ref{sec:appregexmatching}, we extend our results of Section \ref{sec:regex} and  
  give an algorithm, without any assumption on the alphabet size, using $O(\frac{mn\log(d+2)}{k} + n + m\log m)$ time
and $O(2^k + m)$ space, for any $k\leq w$.

\paragraph{String Edit Distance} We conclude by giving a simple way to
improve the complexity of the \textsc{String Edit Distance} problem,
which is defined as that of computing the minimum number of edit
operations needed to transform given string $S$ of length $m$ into
given string $T$ of length $n$.  The standard dynamic programming
solution to this problem uses $O(mn)$ time and $O(\min(m,n))$
space. The fastest algorithm for this problem, due to Masek and
Paterson~\cite{MP1980}, achieves $O(mn/k^2 +m + n)$ time and $O(2^k +
\min(n,m))$ space for any $k \leq w$. However, this algorithm assumes
a constant size alphabet. For long word sizes faster algorithms can be obtained~\cite{Myers1999, BYN1999}. See also the survey by Navarro~\cite{Navarro2001a}.

In Section \ref{sec:stringedit}, we show how to
achieve $O(nm\log^{2} k/k^2 + m + n)$ time and $O(2^k + \min(n,m))$ space
for any $k \leq w$ for an arbitrary alphabet. Hence, we remove the
dependency of the alphabet at the cost of a $\log^{2} k$ factor to the
running time.

\paragraph{Subsequence Indexing}
We also consider a special case of regular expression matching. Given
text $T$, the \textsc{Subsequence Indexing} problem is to preprocess
$T$ to allow queries of the form ``is $Q$ a subsequence of $T$?''
Baeza-Yates~\cite{BaezaYates1991} showed that this problem can be
solved with $O(n)$ preprocessing time and space, and query time
$O(m\log n)$, where $Q$ has length $m$ and $T$ has length $n$.
Conversely, one can achieve queries of time $O(m)$ with $O(n\sigma)$
preprocessing time and space.  As before, $\sigma$ is the size of the
alphabet. 

In Section \ref{sec:subseq}, we give an algorithm that improves the former
results to $O(m\log\log\sigma)$ query time or the latter result to
$O(n\sigma^{\epsilon})$ preprocessing time and space.

\section{Regular Expression Matching}\label{sec:regex} Given an string $Q$ and a
regular expression $R$ the \textsc{Regular Expression Matching}
problem is to determine if $Q$ is in the language given by $R$. Let
$n$ and $m$ be the sizes of $Q$ and $R$, respectively. In this section
we show that \textsc{Regular Expression Matching} can be solved in
$O(mn/k + n + m\log m)$ time and $O(2^{k}+m)$ space, for $k \leq w$.

\subsection{Regular Expressions and NFAs} We briefly review
Thompson's construction and the standard node set simulation.  The set
of \emph{regular expressions} over $\Sigma$ is defined recursively as
follows:
\begin{itemize}
\item A character $\alpha \in \Sigma$ is a regular expression.
\item If $S$ and $T$ are regular expressions then so is the
  \emph{catenation}, $(S)\cdot(T)$, the \emph{union}, $(S)|(T)$, and
  the \emph{star}, $(S)^*$.
\end{itemize}
Unnecessary parentheses can be removed by observing that $\cdot$ and
$|$ are associative and by using the standard precedence of the
operators, that is $*$ precedes $\cdot$, which in turn precedes $|$. Furthermore, we will often remove the $\cdot$ when writing regular expressions.  The \emph{language} $L(R)$ generated by $R$ is the set of all strings
matching $R$. The \emph{parse tree} $T(R)$ of $R$ is the rooted and ordered
tree representing the hierarchical structure of $R$. All leaves are represented by a character in $\Sigma$ and all internal nodes are labeled $\cdot$, $|$, or $^*$. We assume that parse trees are binary and constructed such that they are in one-to-one correspondence with the regular expressions. An example parse tree of the
regular expression $ac|a^*b$ is shown in Fig. \ref{fig:clustering}(a).

A \emph{finite automaton} $A$ is a tuple $A = (G, \Sigma, \theta, \Phi)$ such that,
\begin{itemize}
  \item $G$ is a directed graph,
  \item Each edge $e \in E(G)$ is labeled with a character $\alpha \in \Sigma$ or $\epsilon$,
  \item $\theta \in V(G)$ is a \emph{start node},
  \item $\Phi \subseteq V(G)$ is the set of \emph{accepting nodes}.
\end{itemize} 
$A$ is a \emph{deterministic finite automaton} (DFA) if $A$ does not contain any $\epsilon$-edges, and for each node $v \in V(G)$ all outcoming edges have different labels. Otherwise, $A$ is a \emph{non-deterministic automaton} (NFA). We say that $A$ \emph{accepts} a string $Q$ if there is a path from $\theta$ to a node in $\Phi$ which spells out $Q$. 

\begin{figure}[t] 
  \centering \includegraphics[scale=.5]{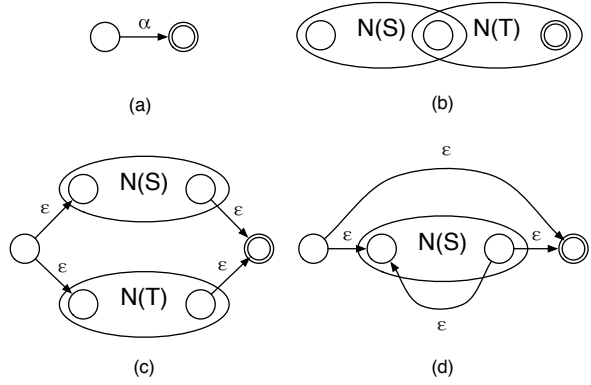}
  \caption{Thompson's NFA construction. The regular expression for a character $\alpha \in \Sigma$ correspond to NFA $(a)$. If $S$ and $T$ are regular expression then $N(ST)$, $N(S|T)$, and
    $N(S^*)$ correspond to NFAs $(a)$, $(b)$, and $(c)$, respectively.
    Accepting nodes are marked with a double circle.}
  \label{fig:thompson}
\end{figure}

Using Thompson's method \cite{Thomp1968} we can recursively construct an NFA $N(R)$ accepting all strings in $L(R)$. The set of rules is presented below and illustrated in
Fig.~\ref{fig:thompson}. 

\begin{itemize}
\item $N(\alpha)$ is the automaton consisting of a start node
  $\theta_\alpha$, accepting node $\phi_\alpha$, and an $\alpha$-edge
  from $\theta_\alpha$ to $\phi_\alpha$.
\item Let $N(S)$ and $N(T)$ be automata for regular expression $S$ and
  $T$ with start and accepting nodes $\theta_S$, $\theta_T$, $\phi_S$,
  and $\phi_T$, respectively. Then, NFAs for $N(S\cdot T)$, $N(S|T)$,
  and $N(S^*)$ are constructed as follows:
  \begin{relate}
  \item[$N(ST)$:] Merge the nodes $\phi_S$ and $\theta_T$ into a
    single node. The new start node is $\theta_S$ and the new
    accepting node is $\phi_T$.
  \item[$N(S|T)$:] Add a new start node $\theta_{S|T}$ and new
    accepting node $\phi_{S|T}$. Then, add $\epsilon$ edges from
    $\theta_{S|T}$ to $\theta_S$ and $\theta_T$, and from $\phi_S$ and
    $\phi_T$ to $\phi_{S|T}$.
  \item [$N(S^*)$:] Add a new start node $\theta_{S^*}$ and new
    accepting node $\phi_{S^*}$. Then, add $\epsilon$ edges from
    $\theta_{S^*}$ to $\theta_S$ and $\phi_{S^*}$, and from $\phi_S$
    to $\phi_{S^*}$ and $\theta_S$.
\end{relate}
\end{itemize}
By construction, $N(R)$ has a single start and accepting node, denoted
$\theta$ and $\phi$, respectively. $\theta$ has no incoming edges and $\phi$ has no outcoming edges. 
The total number of nodes is at most $2m$ and since each node has at most $2$ outgoing edges that the
total number of edges is less than $4m$.  Furthermore, all incoming
edges have the same label, and we denote a node with incoming
$\alpha$-edges an \emph{$\alpha$-node}. Note that the star
construction in Fig. \ref{fig:thompson}(d) introduces an edge from the
accepting node of $N(S)$ to the start node of $N(S)$. All such edges
in $N(R)$ are called \emph{back edges} and all other edges are
\emph{forward edges}. We need the following important property of
$N(R)$.
\begin{lemma}[Myers \cite{Myers1992}]\label{lem:cyclepath}
  Any cycle-free path in $N(R)$ contains at most one back edge.
\end{lemma}
For a string $Q$ of length $n$ the standard node-set simulation of
$N(R)$ on $Q$ produces a sequence of node-sets $S_0, \ldots, S_n$. A node $v$ is in $S_i$ iff there is a path from $\theta$ to $v$ that spells out the $i$th prefix of $Q$.
The simulation can be implemented with the following simple
operations. Let $S$ be a node-set in $N(R)$ and let $\alpha$ be a
character in $\Sigma$.
\begin{relate}
\item[$\Move(S,\alpha)$:] Compute and return the set of nodes
  reachable from $S$ via a single $\alpha$-edge.
\item[$\Close(S)$:] Compute and return the set of nodes reachable from
  $S$ via $0$ or more $\epsilon$-edges.
\end{relate}
The number of nodes and edges in $N(R)$ is $O(m)$, and both operations
are implementable in $O(m)$ time.  The simulation proceed as follows:
Initially, $S_0 := Close(\{\theta\})$. If $Q[j]=\alpha$, $1\leq j \leq
n$, then $S_j := \Close(\Move(S_{j-1}, \alpha))$. Finally, $Q \in
L(R)$ iff $\phi \in S_n$. Since each node-set $S_j$ only depends on
$S_{j-1}$ this algorithm uses $O(mn)$ time $O(m)$ space. 

\subsection{Outline of Algorithm}

Our result is based on a new and more compact encoding of small subautomata used within Myers' algorithm~\cite{Myers1992} supporting constant time $\Move$ and $\Close$ operations. For our purposes and for completeness, we restate Myers' algorithm in Sections~\ref{sec:clustering} and \ref{sec:simulation}, such that the dependency on the $\Move$ and $\Close$ operations on subautomata is exposed. The new encoding is presented in Section~\ref{sec:representation}.

\subsection{Decomposing the NFA}\label{sec:clustering} 
\begin{figure}[ht]
  \centering \includegraphics[scale=.5]{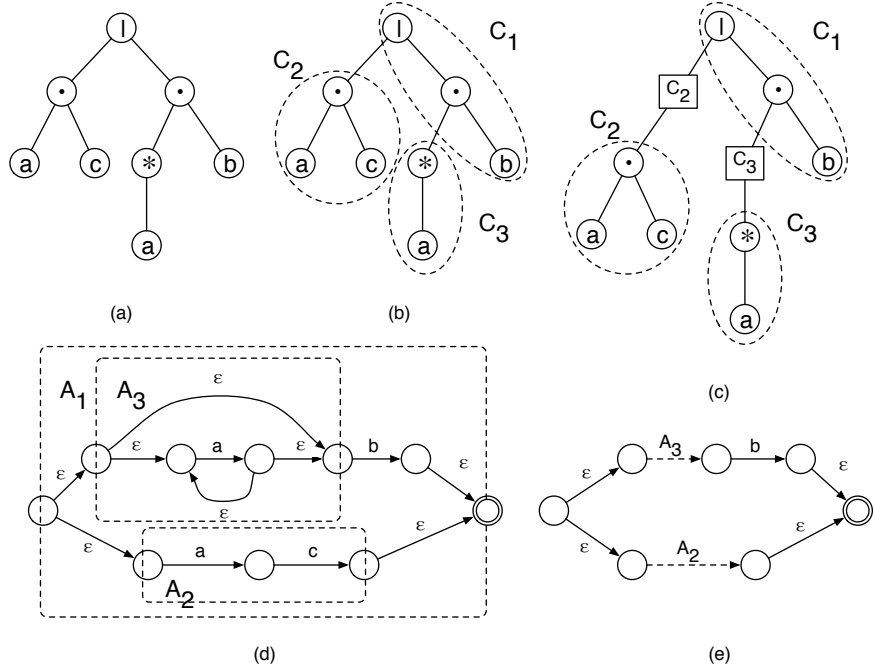}
   \caption{(a) The parse tree for the regular expression
     $ac|a^*b$. (b) A clustering of $(a)$ into
     node-disjoint connected subtrees $C_1$, $C_2$, and $C_3$.  Here,
     $x=3$. (c) The clustering from (b) extended with pseudo-nodes.
     (d) The automaton for the parse tree divided into subautomata
     corresponding to the clustering. (e) The subautomaton $A_1$ with
     pseudo-edges corresponding to the child automata.}
   \label{fig:clustering}
\end{figure}

In this section we show
how to decompose $N(R)$ into small subautomata. In the final algorithm
transitions through these subautomata will be simulated in constant
time.  The decomposition is based on a clustering of the parse tree
$T(R)$. Our decomposition is similar to the one given in
\cite{Myers1992, WMM1995}.  A \emph{cluster} $C$ is a connected
subgraph of $T(R)$. A \emph{cluster partition} $CS$ is a partition of
the nodes of $T(R)$ into node-disjoint clusters. Since $T(R)$ is a
binary tree, a bottom-up procedure yields the following lemma.
\begin{lemma}\label{lem:clustering}
  For any regular expression $R$ of size $m$ and a parameter $x$, it
  is possible to build a cluster partition $CS$ of $T(R)$, such that
  $|CS| = O(m/x)$ and for any $C\in CS$ the number of nodes in $C$ is
  at most $x$.
\end{lemma}
An example clustering of a parse tree is shown in
Fig.~\ref{fig:clustering}(b).

Before proceeding, we need some definitions.  Assume that $CS$ is a
cluster partition of $T(R)$ for a some yet-to-be-determined parameter
$x$. Edges adjacent to two clusters are \emph{external edges} and all
other edges are \emph{internal edges}. Contracting all internal edges
induces a \emph{macro tree}, where each cluster is represented by a
single \emph{macro node}. Let $C_v$ and $C_w$ be two clusters with
corresponding macro nodes $v$ and $w$. We say that $C_v$ is a
\emph{parent cluster} (resp. \emph{child cluster}) of $C_w$ if $v$ is
the parent (resp. child) of $w$ in the macro tree. The \emph{root
  cluster and leaf clusters} are the clusters corresponding to the
root and the leaves of the macro tree.

Next we show how to decompose $N(R)$ into small subautomata. Each
cluster $C$ will correspond to a subautomaton $A$ and we use the terms
child, parent, root, and leaf for subautomata in the same way we do
with clusters. For a cluster $C$, we insert a special
\emph{pseudo-node} $p_i$ for each child cluster $C_1, \ldots, C_\ell$ in
the middle of the external edge connecting $C$ and $C_i$. Now, $C$'s
subautomaton $A$ is the automaton corresponding to the parse tree
induced by the set of nodes $V(C) \cup \{p_1, \ldots, p_\ell\}$. The
pseudo-nodes are alphabet placeholders, since the leaves of a
well-formed parse tree must be characters.

In $A$, child automaton $A_i$ is represented by its start and
accepting node $\theta_{A_i}$ and $\phi_{A_i}$ and a
\emph{pseudo-edge} connecting them. An example of these definitions is
given in Fig.~\ref{fig:clustering}. Any cluster $C$ of size at most
$x$ has less than $2x$ pseudo-children and therefore the size of the
corresponding subautomaton is at most $6x$. Note, therefore, that
automata derived from regular expressions can be thus decomposed into
$O(m/z)$ subautomata each of size at most $z$, by
Lemma~\ref{lem:clustering} and the above construction.

\subsection{Simulating the NFA}\label{sec:simulation}
In this section we show how to do a node-set simulation of $N(R)$
using the subautomata. We compactly represent node-set of each subautomata in a bit string and in the next section we will show how to manipulate these node-set efficiently using a combination of the Four Russian Technique and standard word operations. This approach is often called  \emph{bit-parallelism}~\cite{BYG1992}. 

Recall that each subautomaton has size less
than $z$. Topologically sort all nodes in each subautomaton $A$
ignoring back edges. This can be done for all subautomata in total
$O(m)$ time. We represent the current node-set $S$ of $N(R)$ compactly using a bitvector for each subautomaton. Specifically, for each subautomaton $A$ we store a \emph{characteristic bitvector} $\vec{B} = [b_1,
\ldots, b_z]$, where nodes in $\vec{B}$ are indexed by the their
topological order, such that $\vec{B}[i] = 1$ iff the $i$th node is in $S$. If $A$ contains fewer than $z$ nodes we leave the remaining values undefined. For simplicity, we will refer to the \emph{state} of $A$ as the node-set represented by the characteristic vector stored at $A$. Similarly, the state of $N(R)$ is the set of characteristic vectors representing $S$. The state of a node is the bit indicating if the node is in $S$. Since any child $A'$ of $A$ overlap at the nodes $\theta_{A'}$ and $\phi_{A'}$ we will ensure that the state of $\theta_{A'}$ and $\phi_{A'}$ is the same in the characteristic vectors of both $A$ and $A'$. 


Below we present appropriate move and $\epsilon$-closure operations defined on subautomata. Due to the overlap between parent and child nodes these operations take a bit $b$ which will use to propagate the new state of the start node. For each subautomaton $A$, characteristic vector $\vec{B}$, bit $b$, and character $\alpha \in \Sigma$
define:
\begin{relate}
\item[$\Move^A(\vec{B}, b, \alpha)$:] Compute the state $\vec{B}'$ of all nodes in
  $A$ reachable via a single $\alpha$ edge from $\vec{B}$. If $b=0$, return
  $\vec{B}'$, else return $\vec{B}' \cup \{\theta_A\}$.
\item[$\Close^A(\vec{B}, b)$:] Return the set $\vec{B}'$ of all nodes in $A$
  reachable via a path of $0$ or more $\epsilon$-edges from $\vec{B}$, if
  $b=0$, or reachable from $\vec{B} \cup \{\theta_A\}$, if $b = 1$.
\end{relate}
We will later show how to implement these operations in constant time and
total $2^{O(k)}$ space when $z = \Theta(k)$. Before doing so we show
how to use these operations to perform the node-set simulation of
$N(R)$. Assume that the current node-set of $N(R)$ is represented by
its characteristic vector for each subautomaton. The following $\Move$ and $\Close$ operations recursively traverse the hierarchy of subautomata top-down. At each subautomata the current state of $N(R)$ is modified using primarily  $\Move^A$ and  $\Close^A$. For any subautomaton $A$, bit $b$, and character $\alpha \in \Sigma$ define:
\begin{relate}
\item[$\Move(A, b, \alpha)$:] Let $\vec{B}$ be the current state of $A$
  and let $A_1, \ldots, A_\ell$ be children of $A$ in topological order
  of their start node.
\begin{enumerate}
\item Compute $\vec{B}':= \Move^A(\vec{B}, b, \alpha)$.
\item For each $A_i$, $1 \leq i \leq \ell$, 
\begin{enumerate}
\item Compute $f_i := \Move(A_i, b_i, \alpha)$, where $b_i = 1$ iff $\theta_{A_i} \in \vec{B}'$.
\item If $f_i = 1$ set $\vec{B}' := \vec{B}' \cup \{\phi_{A_i}\}$.
\end{enumerate}
\item Store $\vec{B}'$ and return the value $1$ if $\phi_{A} \in \vec{B}'$ and $0$ otherwise.
\end{enumerate}
\item[$\Close(A, b)$:] Let $\vec{B}$ be the current state of $A$ and let
  $A_1, \ldots, A_\ell$ be children of $A$ in topological order of their
  start node.
\begin{enumerate}
\item Compute $\vec{B}' := \Close^A(\vec{B}, b)$.
\item For each child automaton $A_i$, $1 \leq i \leq \ell$,
\begin{enumerate}
\item Compute $f_i := \Close(A_i, b_i)$, where $b_i = 1$ if
  $\theta_{A_i} \in \vec{B}'$.
\item If $f_i = 1$ set $\vec{B}' := \vec{B}' \cup \{\phi_{A_i}\}$.
\item $\vec{B}' := \Close^A(\vec{B}, b)$.
\end{enumerate}
\item Store $\vec{B}'$ and return the value $1$ if $\phi_{A} \in \vec{B}'$ and
  $0$ otherwise.
\end{enumerate}
\end{relate}
The ``store'' in line 3 of both operations updates the state of the subautomaton. The node-set simulation of $N(R)$ on string $Q$ of length $n$ produces the states $S_0, \ldots, S_n$ as follows.  Let $A_r$ be the root automaton. Initialize the state of $N(R)$ to be empty, i.e., set all bitvectors to $0$. $S_0$ is computed by calling $\Close(A_r, 1)$ twice. Assume that $S_{j-1}$, $1 \leq j \leq n$, is the current state of $N(R)$ and let $\alpha = Q[j]$.  Compute $S_j$ by calling $\Move(A_r, 0, \alpha)$ and then calling $\Close(A_r, 0)$ twice.  Finally, $Q \in L(R)$ iff $\phi \in S_n$.

We argue that the above algorithm is correct. To do this we need to show that the call to the $\Move$ operation and the two calls to the $\Close$ operation simulates the standard $\Move$ and $\Close$ operations.

First consider the $\Move$ operation. Let $S$ be the state of $N(R)$ and let $S'$ be the state after a call to $\Move(A_r, 0, \alpha)$. Consider any subautomaton $A$ and let $\vec{B}$ and $\vec{B}'$ be the bitvectors of $A$ corresponding to states $S$ and $S'$, respectively. We first show by induction that after $\Move(A, 0,  \alpha)$ the new state $\vec{B}'$ is the set of nodes reachable from $\vec{B}$ via a single $\alpha$-edge in $N(R)$. For $\Move(A, 1, \alpha)$ a similar argument shows that new state is the union of the set of nodes reachable from $\vec{B}$ via a single $\alpha$-edge and $\{\theta_A\}$. 

Initially, we compute $\vec{B}' := \Move^A(\vec{B}, 0, \alpha)$. Thus $\vec{B}'$ contains the set of nodes reachable via a single $\alpha$-edge in $A$. If $A$ is a leaf automaton then $\vec{B}'$ satisfies the property and the algorithm returns. Otherwise, there may be an $\alpha$-edge to some accepting node $\phi_{A_i}$ of a child automaton $A_i$. Since this edge is not contained $A$, $\phi_{A_i}$ is not initially in $\vec{B}'$. However, since each child is handled recursively in topological order and the new state of start and accepting nodes are propagated, it follows that $\phi_{A_i}$ is ultimately added to $\vec{B}'$. Note that since a single node can be the accepting node of a child $A_i$ and the start node of child $A_{i+1}$, the topological order is needed to ensure a consistent update of the state.

It now follows that the state $S'$ of $N(R)$ after $\Move(A_r, 0, \alpha)$, consists of all nodes reachable via a single $\alpha$-edge from $S$. Hence, $\Move(A_r, 0, \alpha)$ correctly simulates a standard $\Move$ operation.

Next consider the two calls to the $\Close$ operation. Let $S$ be the state of $N(R)$ and let $S'$ be the state after  the first call to $\Close(A_r,0)$. As above consider any subautomaton $A$ and let $\vec{B}$ and $\vec{B}'$ be the bitvectors of $A$ corresponding to $S$ and $S'$, respectively. We show by induction that after $\Close(A,0)$ the state $\vec{B}'$ \emph{contains} the set of nodes in $N(R)$ reachable via a path of $0$ or more \emph{forward} $\epsilon$-edges from $\vec{B}$. Initially, $\vec{B}' := \Close^A(\vec{B}, 0)$, and hence $\vec{B}'$ contains  all nodes reachable via a path of $0$ or more $\epsilon$-edges from $\vec{B}$, where the path consists solely of edges in $A$. If $A$ is a leaf automaton, the result immediately holds. Otherwise, there may be a path of $\epsilon$-edges to a node $v$  going through the children of $A$. As above, the recursive topological processing of the children ensures that $v$ is added to $\vec{B}'$. 

Hence, after the first call to $\Close(A_r, 0)$ the state $S'$ contains all nodes reachable from $S$ via a path of $0$ or more forward $\epsilon$-edges. By a similar argument it follows that the second call to $\Close(A_r, 0)$ produces the state $S''$ that contains all the nodes reachable from $S$ via a path of $0$ or more forward $\epsilon$-edge and  $1$ back edge. However, by Lemma~\ref{lem:cyclepath} this is exactly the set of nodes reachable via a path of $0$ or more $\epsilon$-edges. Furthermore, since $\Close(A_r, 0)$ never produces a state with nodes that are not reachable through $\epsilon$-edges, it follows that the two calls to $\Close(A_r, 0)$ correctly simulates a standard $\Close$ operation. 

Finally, note that if we start with a state with no nodes, we can compute the state $S_0$ in the node-set simulation by calling $\Close(A_r, 1)$ twice. Hence, the above algorithm correctly solves \textsc{Regular Expression Matching}.

If the subautomata have size at most $z$ and $\Move^A$ and $\Close^A$ can be computed in constant time the above algorithm computes a step in the node-set simulation in $O(m/z)$ time. In the following section
we show how to do this in $O(2^{k})$ space for $z = \Theta(k)$. Note that computing the clustering uses an additional $O(m)$ time and space.

\subsection{Representing Subautomata}\label{sec:representation} To efficiently
represent $\Move^A$ and $\Close^A$ we apply the Four Russian trick.
Consider a straightforward code for $\Move^A$: Precompute the value of
$\Move^A$ for all $\vec{B}$, both values of $b$, and all characters
$\alpha$. Since the number of different bitvectors is $2^{z}$ and the
size of the alphabet is $\sigma$, this table has $2^{z+1}\sigma$
entries.  Each entry can be stored in a single word, so the table also uses
a total of $2^{z+1}\sigma$ space. The total number of subautomata is
$O(m/z)$, and therefore the total size of these tables is an unacceptable
$O(\frac{m}{z} \cdot 2^{z} \sigma)$.  

To improve this we use a more elaborate approach. First we factor out
the dependency on the alphabet, as follows. For all subautomata $A$ and all
characters $\alpha \in \Sigma$ define:
\begin{relate}
\item[$\Succ^A(\vec{B})$:] Return the set of all nodes in $A$ reachable
  from $\vec{B}$ by a single edge.
\item[$\Eq^A(\alpha)$:] Return the set of all $\alpha$-nodes in $A$.
\end{relate}
Since all incoming edges to a node are labeled with the same character
it follows that,
\begin{equation*}
\Move^A(\vec{B}, b, \alpha) =
\begin{cases}
  \Succ^A(\vec{B}) \cap \Eq^A(\alpha) &\text{if $b=0$}, \\
  (\Succ^A(\vec{B}) \cap \Eq^A(\alpha)) \cup \{\theta_A\} & \text{if $b=1$}.
\end{cases} 
\end{equation*}
Hence, given $\Succ^A$ and $\Eq^A$ we can implement $\Move^A$ in
constant time using bit operations. To efficiently represent $\Eq^{A}$,
for each subautomaton $A$, store the value of $\Eq^A(\alpha)$ in a
hash table. Since the total number of different characters in $A$ is
at most $z$ the hash table $\Eq^A$ contains at most $z$ entries.
Hence, we can represent $\Eq^{A}$ for all subautomata is $O(m)$ space and
constant worst-case lookup time. The preprocessing time is $O(m)$
w.h.p.. To get a worst-case preprocessing bound we use the deterministic dictionary
of~\cite{HMP2001} with $O(m \log m)$ worst-case preprocessing time.

We note that the idea of using $\Eq^A(\alpha)$ to represent the $\alpha$-nodes is not new and has been used in several string matching algorithms, for instance, in the classical Shift-Or algorithm~\cite{BYG1992} and in the recent optimized DFA construction for regular expression matching~\cite{NR2004}.

To represent $\Succ$ compactly we proceed as follows. Let $\hat{A}$ be
the automaton obtained by removing the labels from edges in $A$.
$\Succ^{A_1}$ and $\Succ^{A_2}$ compute the same function if
$\hat{A_1} = \hat{A_2}$. Hence, to represent $\Succ$ it suffices to
precompute $\Succ$ on all possible subautomata $\hat{A}$. By the
one-to-one correspondence of parse trees and automata we have that
each subautomata $\hat{A}$ corresponds to a parse tree with leaf
labels removed. Each such parse tree has at most
$x$ internal nodes and $2x$ leaves. The number of rooted, ordered,
binary trees with at most $3x$ nodes is less than $2^{6x+1}$, and for
each such tree each internal node can have one of $3$ different labels.
Hence, the total number of distinct subautomata is less than $2^{6x +
  1}3^x$. Each subautomaton has at most $6x$ nodes and therefore the
result of $\Succ^A$ has to be computed for each of the $2^{6x}$
different values for $\vec{B}$ using $O(x2^{6x})$ time. Therefore we can
precompute all values of $\Succ$ in $O(x2^{12x + 1}3^x)$ time.
Choosing $x$ such that $x+\frac{\log x}{12 + \log 3} \leq \frac{k -
  1}{12 + \log 3}$ gives us $O(2^k)$ space and preprocessing time.

Using an analogous argument, it follows that $\Close^A$ can be
precomputed for all distinct subautomata within the same complexity.
By our discussion in the previous sections and since $x = \Theta(k)$ we have shown the following theorem:
\begin{theorem}
  For regular expression $R$ of length $m$, string $Q$ of length
  $n$, and $k \leq w$,  \textsc{Regular Expression Matching} can be solved in $O(mn/k +  n + m\log m)$ time and $O(2^k + m)$ space.
\end{theorem}

\section{Approximate Regular Expression Matching}\label{sec:appregexmatching}
Given a string $Q$, a regular expression $R$, and an integer $d \geq
0$, the \textsc{Approximate Regular Expression Matching} problem is to
determine if $Q$ is within edit distance $d$ to a string in
$L(R)$. In this section we extend our solution for \textsc{Regular Expression
  Matching} to \textsc{Approximate Regular Expression Matching}. 
Specifically, we show that the problem can be solved in $O(\frac{mn\log(d+2)}{k} + n + m\log m)$ time and $O(2^k +m)$ space, for any $k\leq w$. 
 
Our result is achieved through a new encoding of subautomata within an algorithm by Wu et al.~\cite{WMM1995} in a style similar to the above result for \textsc{Regular Expression Matching}. For completeness we restate the algorithm of Wu et al.~\cite{WMM1995} in Sections~\ref{sec:dynamicprogram} and \ref{sec:simulrecurrence}. The new encoding is given in Section~\ref{sec:representingappx}.

\subsection{Dynamic Programming Recurrence}\label{sec:dynamicprogram} Our algorithm is
based on a dynamic programming recurrence due to Myers and
Miller~\cite{MM1989}, which we describe below. Let $\Delta(v,i)$
denote the minimum over all paths ${\cal P}$ between $\theta$ and $v$
of the edit distance between ${\cal P}$ and the $i$th prefix of $Q$. The recurrence avoids cyclic dependencies from the back
edges by splitting the recurrence into two passes. Intuitively, the
first pass handles forward edges and the second pass propagates values
from back edges.  The \emph{pass-1 value} of $v$ is denoted
$\Delta_1(v,i)$, and the \emph{pass-2 value} is $\Delta_2(v,i)$. For a
given $i$, the \emph{pass-1 (resp. pass-2) value of $N(R)$} is the set of
pass-1 (resp. pass-2) values of all nodes of $N(R)$.  For all
$v$ and $i$, we set $\Delta(v,i) = \Delta_2(v,i)$.

The set of \emph{predecessors} of $v$ is the set of nodes $\Pre(v) =
\{w \mid \text{ $(w, v)$ is an edge}\}$.  We define
$\overline{\Pre}(v) = \{w \mid \text{ $(w,v)$ is a forward edge}\}$.
For notational convenience, we extend the definitions of $\Delta_1$
and $\Delta_2$ to apply to sets, as follows: $\Delta_1(\Pre(v),i) =
\min_{w\in \Pre(v)} \Delta_1(w, i)$ and
$\Delta_1(\overline{\Pre}(v),i) = \min_{w\in \overline{\Pre}(v)}
\Delta_1(w, i)$, and analogously for $\Delta_2$. The pass-1 and pass-2
values satisfy the following recurrence:
 
\begin{align*}
  \Delta_2(\theta, i) &= \Delta_1(\theta, i) = i \qquad \text{$0 \leq i \leq n$}. \\
  \Delta_2(v,0) &= \Delta_1(v, 0) = \min
\begin{cases}
  \Delta_2(\overline{\Pre}(v), 0) + 1 & \text{if $v$ is a $\Sigma$-node}, \\
  \Delta_2(\overline{\Pre}(v), 0) & \text{if $v \neq \theta$ is an
    $\epsilon$-node}.
\end{cases}  \\
\intertext{$\qquad$ For $1\leq i \leq n$,}
\Delta_1(v, i) &=
\begin{cases}
  \min(\Delta_2(v, i-1) + 1, \Delta_2(\Pre(v), i) + \lambda(v,Q[i]),
  \Delta_1(\overline{\Pre}(v), i) + 1)&
  \text{if $v$ is a $\Sigma$-node}, \\
  \Delta_1(\overline{\Pre}(v), i) & \text{if $v\neq \theta$ is an $\epsilon$-node},
\end{cases} \\
\intertext{$\qquad$ where $\lambda(v,Q[i]) = 1$ if $v$ is a
  $Q[i]$-node and $0$ otherwise,}  
  \Delta_2(v,i) &=
\begin{cases}
  \min(\Delta_1(\Pre(v), i), \Delta_2(\overline{\Pre}(v), i)) + 1
  & \text{if $v$ is a $\Sigma$-node}, \\
  \min(\Delta_1(\Pre(v), i), \Delta_2(\overline{\Pre}(v), i))
  & \text{if $v$ is a $\epsilon$-node}. \\
\end{cases} 
\end{align*}
A full proof of the correctness of the above recurrence can be found
in~\cite{MM1989, WMM1995}. Intuitively, the first pass handles forward
edges as follows: For $\Sigma$-nodes the recurrence handles
insertions, substitution/matches, and deletions (in this order). For
$\epsilon$-nodes the values computed so far are propagated.
Subsequently, the second pass handles the back edges. For our problem
we want to determine if $Q$ is within edit distance $d$. Hence, we can
replace all values exceeding $d$ by $d+1$.  

\subsection{Simulating the Recurrence}\label{sec:simulrecurrence} Our algorithm now
proceeds analogously to the case with $d=0$ above.  We will decompose
the automaton into subautomata, and we will compute the above dynamic
program on an appropriate encoding of the subautomata, leading to a
small-space speedup.

As before, we decompose $N(R)$ into subautomata of size less than $z$.
For a subautomaton $A$ we define operations $\Next^A_1$ and
$\Next^A_2$ which we use to compute the pass-1 and pass-2 values of
$A$, respectively. However, the new (pass-1 or pass-2) value of $A$
depends on pseudo-edges in a more complicated way than before: If $A'$
is a child of $A$, then all nodes preceding $\phi_{A'}$ depend on
the value of $\phi_{A'}$. Hence, we need the value of $\phi_{A'}$
before we can compute values of the nodes preceding $\phi_{A'}$. To
address this problem we partition the nodes of a subautomaton as
described below.

For each subautomaton $A$ topologically sort the nodes (ignoring back
edges) with the requirement that for each child $A'$ the start and
accepting nodes $\theta_{A'}$ and $\phi_{A'}$ are consecutive in the
order. Contracting all pseudo-edges in $A$ this can be done for all
subautomata in $O(m)$ time. Let $A_1, \ldots, A_\ell$ be the children of
$A$ in this order. We partition the nodes in $A$, except $\{\theta_A\}
\cup \{\phi_{A_1}, \ldots, \phi_{A_\ell}\}$ , into $\ell+1$ \emph{chunks}.
The first chunk is the nodes in the interval $[\theta_A + 1,
\theta_{A_1}]$. If we let $\phi_{A_{\ell+1}} = \phi_A$, then the $i$th
chunk, $1\leq i \leq \ell+1$, is the set of nodes in the interval
$[\phi_{A_{i-1}}+1, \theta_{A_i}]$. A leaf automaton has a single
chunk consisting of all nodes except the start node. We represent the
$i$th chunk in $A$ by a characteristic vector $\vec{L_i}$ identifying
the nodes in the chunks, that is, $\vec{L_i}[j] = 1$ if node $j$ is in the $i$th chunk and $0$ otherwise. From the topological order we can compute all chunks and their corresponding characteristic vectors in total $O(m)$
time.

The value of $A$ is represented by a vector $\vec{B} = [b_1, \ldots,
b_z]$, such that $b_i \in [0, d+1]$. Hence, the total number of bits
used to encode $\vec{B}$ is $z\ceil{\log d+2}$ bits.  For an
automaton $A$, characteristic vectors $\vec{B}$ and $\vec{L}$, and a
character $\alpha \in \Sigma$ define the operations
$\Next^A_1(\vec{B}, \vec{L}, b, \alpha)$ and
$\Next^A_2(\vec{B},\vec{L}, b)$ as the vectors $\vec{B}_1$ and
$\vec{B}_2$, respectively, given by:
\begin{align*}
  \vec{B}_1[v] &= B[v]  \qquad\quad \text{if $v \not\in \vec{L}$} \\
  \vec{B}_1[v] &=
\begin{cases}
  \min(\vec{B}[v] + 1, \vec{B}[\Pre(v)] + \lambda(v, \alpha),  \vec{B}_1[\overline{\Pre}(v)]  + 1) & \text{if $v \in\vec{L}$ is a $\Sigma$-node}, \\
  \vec{B}_1[\Pre(v)] & \text{if $v \in \vec{L}$ is an $\epsilon$-node}
\end{cases} \\
\vec{B}_2[v] &= B[v] \qquad\quad \text{if $v \not\in \vec{L}$}  \\
\vec{B}_2[v] &=
\begin{cases}
  \min(\vec{B}[\Pre(v)], \vec{B}_2[\overline{\Pre}(v)] + 1) & \text{if $v \in\vec{L}$ is a $\Sigma$-node}, \\
  \min(\vec{B}[\Pre(v)], \vec{B}_2[\overline{\Pre}(v)]) & \text{if $v
    \not\in\vec{L}$ is an $\epsilon$-node}
\end{cases} \\
\end{align*}
Importantly, note that the operations only affect the nodes in the
chunk specified by $\vec{L}$. We will use this below to
compute new values of $A$ by advancing one chunk at each step. We use
the following recursive operations. For subautomaton $A$,  integer
$b$, and character $\alpha$ define:
\begin{relate}
\item[$\Next_1(A, b, \alpha)$:] Let $\vec{B}$ be the current value of
  $A$ and let $A_1, \ldots, A_\ell$ be children of $A$ in topological
  order of their start node.
\begin{enumerate}
\item Set $\vec{B}_1 := \vec{B}$ and $\vec{B}_1[\theta_{A}] := b$.
\item For each chunk $L_i$, $1 \leq i \leq \ell$,
        \begin{enumerate}
        \item Compute $\vec{B}_1 := \Next^{A}_1(\vec{B}_1, \vec{L_i},
          \alpha)$.
        \item Compute $f_i := \Next_1(A_i, \vec{B}_1[\theta_{A_i}],
          \alpha)$.
        \item Set $\vec{B}_1[\phi_{A_i}] := f_i$.
\end{enumerate}
\item Compute $\vec{B}_1 := \Next^{A}_1(\vec{B}_1, \vec{L}_{\ell+1},
  \alpha)$.
\item Return $\vec{B}_1[\phi_A]$.
\end{enumerate}
\item[$\Next_2(A, b)$:] Let $\vec{B}$ be the current value of $A$ and
  let $A_1, \ldots, A_\ell$ be children of $A$ in topological order of
  their start node.
\begin{enumerate}
\item Set $\vec{B}_2 := \vec{B}$ and $\vec{B}_2[\theta_{A}] := b$.
\item For each chunk $L_i$, $1 \leq i \leq \ell$,
        \begin{enumerate}
        \item Compute $\vec{B}_2 := \Next^{A}_2(\vec{B}_2,
          \vec{L_i})$.
        \item Compute $f_i := \Next_2(A_i, \vec{B}_2[\theta_{A_i}])$.
        \item Set $\vec{B}_2[\phi_{A_i}] := f_i$.
\end{enumerate}
\item Compute $\vec{B}_2 := \Next^{A}_2(\vec{B}_2, \vec{L}_{\ell+1})$.
\item Return $\vec{B}_2[\phi_A]$.
\end{enumerate}
\end{relate}
The simulation of the dynamic programming recurrence on a string $Q$
of length $n$ proceeds as follows: First encode the initial values of
the all nodes in $N(R)$ using the recurrence. Let $A_r$ be the root
automaton, let $S_{j-1}$ be the current value of $N(R)$, and let
$\alpha = Q[j]$. Compute the next value $S_j$ by calling $\Next_1(A_r,
j, \alpha)$ and then $\Next_2(A_r, j, \alpha)$. Finally, if the
value of $\phi$ in the pass-2 value of $S_n$ is less than $d$, report a match. 

To see the correctness, we need to show that the calls $\Next_1$ and $\Next_2$ operations correctly compute the pass-1 and pass-2 values of $N(R)$. First consider $\Next_1$, and let $A$ be any subautomaton. The key property is that if $p_1$ is the pass-1 value of $\theta_A$ then after a call to $\Next_1(A, p_1, \alpha)$, the value of $A$ is correctly updated to the pass-1 value. This follows by a straightforward induction similar to the exact case. Since the pass-1 value of $\theta$ after reading the $j$th prefix of $Q$ is $j$, the correctness of the call to $\Next_1$ follows. For $\Next_2$ the result follows by an analogous argument.

\subsection{Representing Subautomata}\label{sec:representingappx}
Next we show how to efficiently represent $\Next^A_1$ and $\Next^A_2$.
First consider $\Next^A_1$. Note that again the alphabet size is a
problem. Since the $\vec{B}_1$ value of a node in $A$ depends on other
$\vec{B}_1$ values in $A$ we cannot ``split'' the computation of
$\Next^A_1$ as before. However, the alphabet character only affects
the value of $\lambda(v,\alpha)$, which is $1$ if $v$ is an
$\alpha$-node and $0$ otherwise. Hence, we can represent $\lambda(v,
\alpha)$ for all nodes in $A$ with $\Eq^A(\alpha)$ from the previous
section.  Recall that $\Eq^A(\alpha)$ can be represented for all
subautomata in total $O(m)$ space. With this representation the total
number of possible inputs to $\Next^A_1$ can be represented using
$(d+2)^{z} + 2^{2z}$ bits. Note that for $z = \frac{k}{\log(d+2)}$ we have
that $(d+2)^z = 2^k$.  Furthermore, since $\Next^A_1$ is now alphabet
independent we can apply the same trick as before and only precompute
it for all possible parse trees with leaf labels removed. It follows
that we can choose $z = \Theta(\frac{k}{\log(d+2})$ such that $\Next^A_1$
can precomputed in total $O(2^k)$ time and space. An analogous argument
applies to $\Next^A_2$.  Hence, by our discussion in the previous
sections we have shown that,
\begin{theorem}
  For regular expression $R$ of length $m$, string $Q$ of length $n$,
  and integer $d \geq 0$ \textsc{Approximate Regular Expression
    Matching} can be solved in $O(\frac{mn \log(d+2)}{k} + n + m\log m)$ time and $O(2^k+m)$
  space, for any $k\leq w$.
\end{theorem}

\section{String Edit Distance}\label{sec:stringedit}
The \textsc{String Edit Distance} problem is to compute the minimum number of edit operations needed to transform a string $S$ into a string $T$. Let $m$ and $n$ be the size of $S$ and $T$, respectively. The classical solution to this problem, due to Wagner and Fischer~\cite{WF1974}, fills in the entries of an $m + 1 \times n+1$ matrix $D$. The entry $D_{i,j}$ is the edit distance between $S[1..i]$ and $T[1..j]$, and can be computed using the following recursion:
\begin{align*}
    D_{i,0} &= i   \\
    D_{0,j} &= j  \\
    D_{i,j} &= \min\{D_{i-1, j-1} + \lambda(i,j), D_{i-1, j} + 1, D_{i, j-1} + 1\}
\end{align*}
where $\lambda(i,j) = 0$ if $S[i] = T[j]$ and $1$ otherwise. The edit distance between $S$ and $T$ is the entry $D_{m,n}$. Using dynamic programming the problem can be solved in $O(mn)$ time. When filling out the matrix we only need to store the previous row or column and hence the space used is $O(\min(m,n))$. For further details, see the book by Gusfield~\cite[Chap. 11]{Gusfield1997}. 

The best algorithm for this problem, due to Masek and Paterson~\cite{MP1980}, improves the time to $O(\frac{mn}{k^{2}} +m+ n)$ time and $O(2^k + \min(m,n))$ space, for any $k \leq w$. This algorithm, however, assumes that the alphabet size is constant. In this section we give an algorithm using $O(\frac{mn\log^{2} k}{k^{2}} +m+ n)$ time and $O(2^k + \min(m,n))$ space, for any $k \leq w$, that works for any alphabet. Hence, we remove the dependency of the alphabet at the cost of a $\log^{2} k$ factor.

We first describe the algorithm by Masek and Paterson~\cite{MP1980}, and then modify it to handle arbitrary alphabets. The algorithm uses the Four Russian Technique. The matrix $D$ is divided into \emph{cells} of size $x \times x$ and all possible inputs of a cell is then precomputed and stored in a table. From the above recursion it follows that the values inside each cell $C$ depend on the corresponding substrings in $S$ and $T$, denoted $S_{C}$ and $T_{C}$, and on the values in the top row and the leftmost column in $C$. The number of different strings of length $x$ is $\sigma^{x}$ and hence there are $\sigma^{2x}$ possible choices for $S_{C}$ and $T_{C}$. Masek and Paterson~\cite{MP1980} showed that adjacent entries in $D$ differ by at most one, and therefore if we know the value of an entry there are exactly three choices for each adjacent entry. Since there are at most $m$ different values for the top left corner of a cell it follows that the number of different inputs for the top row and the leftmost column is $m3^{2x}$. In total, there are at $m(\sigma3)^{2x}$ different inputs to a cell. Assuming that the alphabet has constant size, we can choose $x = \Theta(k)$ such that all cells can be precomputed in $O(2^k)$ time and space. The input of each cell is stored in a single machine word and therefore all values in a cell can be computed in constant time. The total number of cells in the matrix is $O(\frac{mn}{k^{2}})$ and hence this implies an algorithm using $O(\frac{mn}{k^{2}} +m +n)$ time and $O(2^k + \min(m,n))$ space. 

We show how to generalize this to arbitrary alphabets. The first observation, similar to the idea in Section~\ref{sec:appregexmatching}, is that the values inside a cell $C$ does not depend on the actual characters of $S_{C}$ and $T_{C}$, but only on the $\lambda$ function on $S_C$ and $T_C$. Hence, we only need to encode whether or not $S_{C}[i] = T_{C}[j]$ for all $1 \leq i,j \leq x$. To do this we assign a code $c(\alpha)$ to each character $\alpha$ that appears in $T_C$ or $S_C$ as follows. If $\alpha$ only appears in only one of $S_C$ or $T_C$ then $c(\alpha) = 0$.  Otherwise, $c(\alpha)$ is the rank of $\alpha$ in the sorted list of characters that appears in both $S_C$ and $T_C$. The representation is given by two vectors $\vec{S}_{C}$ and $\vec{T}_{C}$ of size $x$, where $\vec{S}_C[i] = c(S_C[i])$ and $\vec{T}_C[i] = c(T_C[i])$, for all $i$, $1\leq i \leq x$.  Clearly, $S_C[i] = T_C[j]$ iff $\vec{S}_{C}[i] = \vec{T}_{C}[j]$ and $\vec{S}_{C}[i] > 0$ and $\vec{T}_{C}[j] >0$ and hence $\vec{S}_C$ and $\vec{T}_C$ suffices to represent $\lambda$ on $C$. 

The number of characters appearing in both $T_{C}$ and $S_{C}$ is at most $x$ and hence each entry of the vectors is assigned an integer value in the range $[1, x]$. Thus, the total number of bits needed for both vectors is $2x\ceil{\log x + 1}$. Hence, we can choose $x = \Theta(\frac{k}{\log k})$ such that the input vectors for a cell can be represented in a single machine word. The total number of cells becomes $O(\frac{mn}{x^{2}}) = O(\frac{nm\log^{2} k}{k^{2}})$. Hence, if the input vectors for each cell is available we can use the Four Russian Technique to get an algorithm for \textsc{String Edit Distance} using $O(\frac{nm\log^{2} k}{k^{2}} + m + n)$ time and $O(2^k + \min(m,n))$ space as desired.

Next we show how to compute vectors efficiently. Given any cell $C$, we can identify the characters appearing in both $S_C$ and $T_C$ by sorting $S_C$ and then for each index $i$ in $T_C$ use a binary search to see if $T_C[i]$ appears in $S_C$. Next we sort the characters appearing in both substrings and insert their ranks into the corresponding positions in $\vec{S}_C$ and $\vec{T}_C$. All other positions in the vectors are given the value $0$.  This algorithm uses $O(x\log x)$ time for each cell. However, since the number of cells is $O(\frac{nm}{x^{2}})$ the total time becomes $O(\frac{nm\log x}{x})$, which for our choice of $x$ is $O(\frac{nm (\log k)^{2}}{k})$. To improve this we group the cells into \emph{macro cells} of $y \times y$ cells. We then compute the vector representation for each of these macro cells. The vector representation for a cell $C$ is now the corresponding subvectors of the macro cell containing $C$. Hence, each vector entry is now in the range $[0, \ldots, xy]$ and thus uses $\ceil{\log(xy + 1)}$ bits. Computing the vector representation uses $O(xy \log (xy))$ time for each macro cell and since the number of macro cells is $O(\frac{nm}{(xy)^{2}})$ the total time to compute it is $O(\frac{nm\log(xy)}{xy} +m +n)$. It follows that we can choose $y = k\log k$ and $x = \Theta(\frac{k}{\log k})$ such that vectors for a cell can be represented in a single word. With this choice of $x$ and $y$ we have that $xy = \Theta(k^{2})$ and hence all vectors are computed in $O(\frac{nm\log (xy)}{xy} + m + n) = O(\frac{nm\log k}{k^{2}} + m + n)$ time. Computing the distance matrix dominates the total running time and hence we have shown:
\begin{theorem}
For strings $S$ and $T$ of length $n$ and $m$, respectively, \textsc{String Edit Distance} can be solved in $O(\frac{mn\log^{2} k}{k^{2}} + m + n)$ time and $O(2^k + \min(m,n))$ space.
\end{theorem}

\section{Subsequence Indexing}\label{sec:subseq}
The \textsc{Subsequence Indexing} problem is to preprocess a string $T$ to build a data structure supporting queries of the form:``is $Q$ a subsequence of $T$?'' for any string $Q$. This problem was considered by Baeza-Yates~\cite{BaezaYates1991} who showed the trade-offs listed in Table~\ref{subseqcomplexity}.  We assume throughout the section that $T$ and $Q$ have  $n$ and $m$, respectively. For properties of automata accepting subsequences of string and generalizations of the problem see the recent survey~\cite{CMT2003}.
\begin{table}[t]
  \centering 
  \begin{tabular}{|c|c|c|}
\hline
Space   			& Preprocessing 	& Query \\ \hline
$O(n\sigma)$  		& $O(n\sigma)$ 		& $O(m)$ \\\hline
$O(n\log \sigma)$  	& $O(n \log \sigma)$& $O(m\log \sigma)$ \\ \hline
$O(n)$				& $O(n)$			& $O(m\log n)$ \\ \hline
\end{tabular}
  \caption{Trade-offs for \textsc{Subsequence Indexing}.} \label{subseqcomplexity}
\end{table}

Using recent data structures and a few observations we improve all previous bounds. As a notational shorthand, we
will say that a data structure with preprocessing time and space
$f(n,\sigma)$ and query time $g(m,n,\sigma)$ has complexity
\pqtime{f(n,\sigma)}{g(m,n,\sigma)}

Let us consider the simplest algorithm for \textsc{Subsequence
  Indexing}. One can build a DFA of size $O(n\sigma)$
for recognizing all subsequences of $T$.  To do so, create an
accepting node for each character of $T$, and for node $v_i$,
  corresponding to character $T[i]$, create an edge to
  $v_j$ on character $\alpha$ if $T[j]$ is the first $\alpha$ after
  position $i$.  The start node has edges to the first
  occurrence of each character.  Such an automaton yields an algorithm
  with complexity \pqtime{O(n\sigma)}{O(m)}.
  
  An alternative is to build, for each character $\alpha$, a data
  structure $D_{\alpha}$ with the positions of $\alpha$ in $T$.
  $D_{\alpha}$ should support fast successor queries.  The
  $D_{\alpha}$'s can all be built in a total of linear time and space
  using, for instance, van Emde Boas trees and perfect hashing ~\cite{Boas1977, BKZ1977, MN1990}.  These
  trees have query time $O(\log\log n)$.  We use these vEB trees to simulate the above automaton-based algorithm: whenever we are in state $v_{i}$, and the next character to be read from $P$ is $\alpha$, we look up the successor of $i$ in   $D_{\alpha}$ in $O(\log\log n)$ time.  The complexity of this algorithm is \pqtime{O(n)}{O(m\log\log n}.
  
  We combine these two data structures as follows: Consider an automaton consisting of nodes $u_1,   \ldots, u_{n/\sigma}$, where node $u_i$ corresponds to characters   $T[\sigma(i-1), \ldots, \sigma i - 1]$, that is, each node $u_i$
  corresponds to $\sigma$ nodes in $T$. Within each such node, apply
  the vEB based data structure.  Between such nodes, apply the full
  automaton data structure.  That is, for node $w_i$, compute the
  first occurrence of each character $\alpha$ after $T[\sigma i -1]$.
  Call these \emph{long jumps}.  A edge takes you to a node
  $u_j$, and as many characters of $P$ are consumed with $u_j$ as
  possible.  When no valid edge is possible within $w_j$, take a
  long jump. The automaton uses $O(\frac{n}{\sigma} \cdot \sigma) = O(n)$ space and preprocessing time. The total size of the vEB data structures is $O(n)$. Since each $u_{i}$ consist of at most $\sigma$ nodes, the query time is improved to $O(\log \log \sigma)$. Hence, the complexity of this algorithm is \pqtime{O(n)}{O(m\log \log \sigma)}. To get a trade-off we can replace the vEB data structures by a recent data structure of Thorup~\cite[Thm. 2]{Thorup2003}. This data structure supports successor queries of $x$ integers in the range $[1,X]$ using $O(xX^{1/2^\ell})$ preprocessing time and space with query time $O(\ell+1)$, for $0\leq \ell \leq \log \log X$. Since each of the $n/\sigma$ groups of nodes contain at most $\sigma$ nodes, this implies the following result:
    
\begin{theorem}
  \textsc{Subsequence Indexing} can be solved in
  \pqtime{O(n\sigma^{1/2^\ell})}{O(m(\ell+1))}, for $0\leq \ell \leq\log\log\sigma$.
\end{theorem}

\begin{corollary} 
  \textsc{Subsequence Indexing} can be solved in
  \pqtime{O(n\sigma^{\epsilon})}{O(m)} or \pqtime{O(n)}{O(m\log\log\sigma)}.
\end{corollary}
\begin{proof} We set $\ell$ to be a constant or $\log\log\sigma$, respectively.\qed
\end{proof}

We note that using a recent data structure for rank and select queries on large alphabets by Golynski et al.~\cite{GMR2006} we can also immediately obtain an algorithm using time $O(m\log\log\sigma)$ and space $n\log \sigma + o(n\log \sigma)$ \emph{bits}. Hence, this result matches our fastest algorithm while improving the space from $O(n)$ words to the number of bits needed to store $T$.

\section{Acknowledgments}
Thanks to the anonymous reviewers for many detailed and insightful comments.

\bibliographystyle{abbrv}
\bibliography{./paper}

\end{document}